\documentclass[twocolumn,aps,pra,showpacs,superscriptaddress,floatfix]
{revtex4}
\usepackage[dvips]{graphicx}
\newcommand{\beq}{\begin{equation}}
\newcommand{\eeq}{\end{equation}}
\newcommand{\bea}{\begin{eqnarray}}
\newcommand{\eea}{\end{eqnarray}}
\newcommand{\rf}[1]{(\ref{#1})}

\newcommand{\ket}[1]{\left | #1\right \rangle}
\newcommand{\bra}[1]{\left \langle #1 \right |}

\newcommand{\proj}[1]{\ket{#1} \bra{#1}}
\newcommand{\tr}{{\rm \, Tr}\, }
\newcommand{\affB}{%
     ERATO, Japan Science and Technology Agency}
\newcommand{\affA}{%
     Imai Quantum Computing and Information Project,
     Bunkyo-ku, Tokyo 113-0033, Japan}
\newcommand{\affD}{%
     Communications Research Laboratory,
     Koganei, Tokyo 184-8795, Japan}
\newcommand{\affE}{%
     CREST, Japan Science and Technology Agency}

\begin{document}
\title{Geometrical Conditions for CPTP Maps and
their Application to a Quantum Repeater and a
State-dependent Quantum Cloning Machine}
\author{A. Carlini}
\affiliation{\affA}
\affiliation{\affB}
\author{M. Sasaki}
\affiliation{\affD}
\affiliation{\affE}
\email{e-mail: carlini@qci.jst.go.jp;psasaki@crl.go.jp}

\begin{abstract}

We address the problem of finding optimal CPTP
(completely positive, trace preserving) maps
between a set of binary pure states and another set of
binary generic mixed state in a two dimensional space.
The necessary and sufficient conditions for the existence 
of such CPTP maps can be discussed within a simple geometrical picture. 
We exploit this analysis to show the existence of
an optimal quantum repeater which is superior to the known
repeating strategies for a set of coherent states sent through
a lossy quantum channel.
We also show that the geometrical formulation of the CPTP mapping 
conditions can be a simpler method
to derive a state-dependent quantum (anti) cloning machine
than the study so far based on the explicit solution of
several constraints imposed by unitarity in an extended Hilbert space.

\end{abstract}

\pacs{PACS numbers:03.67.-a, 03.65.Bz, 89.70.+c}
\maketitle

\section{Introduction}\label{introduction}

Suppose that
we receive a quantum state which is drawn from
a parametrized set $\{\hat f_i\}$
with known {\it a priori} probabilities, $\{p_i \}$,
and that
we have another set of states $\{\hat g_i\}$,
which we call {\it templates}, at our disposal.
Our task is
to output an appropriate state function of the templates
that best matches the input.
The meaning of {\it best matching } depends on the task
that we are going to pursue.
For example, we may consider 
an eavesdropping strategy in a quantum cryptosystem,
an action of a quantum repeater in a communication channel,
a state-dependent cloning process, and so on.

The best matching process is generally described by 
a completely positive trace preserving (CPTP) map
from the input to the output state sets.
Unfortunately, however,
the problem of finding the optimal CPTP mapping
between given sets of quantum states is still poorly understood.
For example,
the necessary and sufficient conditions for the existence of
a CPTP mapping between generic mixed states are known only for
binary sets of states in a two dimensional space,
$\{\hat f_1, \hat f_2\}$ and $\{\hat g_1, \hat g_2\}$
\cite{albertiuhlmann} (with $\hat g_i\equiv [\hat I+ \vec{g_i}\cdot 
\hat \sigma]/2$ and, without lack of generality, 
$\vec{g_1}^2=\vec{g_2}^2=g^2$, and $g\in [0, 1]$).
This result has never been exploited for practical purposes of 
quantum information processing. 

In this paper,
we derive a simple geometrical framework for the general
theorem on the existence of CPTP mappings, 
and then apply it to the problem of designing
a quantum optimal repeater
for relaying classical information over a lossy quantum channel,
and to describe a special kind of state-dependent quantum cloning machine. 

Let us suppose that
we are at an intermediate station
and
receive very weak coherent states
$\hat f_1=\proj{\alpha}$
and
$\hat f_2=\proj{-\alpha}$
and that
we must replace these weak signals with stronger ones
consisting of the templates
$\hat g_1=\proj{\beta}$
and
$\hat g_2=\proj{-\beta}$
(where the strict inequality $\vert\beta\vert>\vert\alpha\vert$ holds)
to improve the transmission performance
through the second channel 
which is assumed to be lossy.

We consider CPTP mappings from the inputs
to not only the given template elements
but also a classical mixture of them.
This setting is especially motivated by
a practical scenario 
where
one should find appropriate repeating states 
for the second lossy channel 
and design the optimal mapping for outputting those states.
Actually,
such states will be more or less semi-classical ones 
based on Gaussian states 
because
there will be no much merit to use any non-classical states
for a long-haul lossy channel,
as non-classical states will decohere rapidly and
result in semi-classical ones.
What remains in practice is then
to find an appropriate mixture of coherent state templates.
Thus, we are to design the optimal CPTP map
acting on the input $\hat f_i$, that outputs a quantum state
$\hat \rho_i$ of the form
\beq
\hat f_i \mapsto \hat\rho_i = \sum_j p_{ij} \hat g_j.
\eeq

Another ansatz is then that of quantum cloning.
We are concerned with the special case where,
given $N$ identical inputs $\hat f_i^{\otimes N}$,
we are only able to construct
outputs which are classical mixtures of the templates
consisting of $M$ copies $\hat g_i^{\otimes M}$.
This is a more restricted model than the ones
studied in the literature to date.
However, as seen in section IV, our model provides
a reasonable cloning performance compared with that of 
more general models known so far.
In particular, when one considers the use of quantum cloning
for a lossy quantum channel based on Gaussian states,
our model can be a good practical scenario as mentioned
in the previous paragraph.
An advantage of our method is that we just have to maximize
the chosen figure of merit along a certain curve specifying
the boundary of the allowed CPTP mappings,
unlike the conventional methods that
rely on dealing with all the inequalities for the constraints
imposed by unitarity over extended Hilbert spaces with ancilla.

\section{CPTP Mapping Existence Condition}\label{CPTP mapping cond}

The necessary and sufficient conditions for the existence of
a CPTP mapping between the sets of 2-dim states
derived by Alberti and Uhlmann
\cite{albertiuhlmann}
are expressed in the form
\beq
d_{tr}(\hat f_1, t\hat f_2)\geq
d_{tr}(\hat \rho_1, t\hat \rho_2)~~~;~~~
\forall t\in {\cal{R}^+},
\label{au}
\eeq
where the trace norm distance between two operators
$\hat A$ and $\hat B$
is defined as
$d_{tr}(\hat A, \hat B)
\equiv
\tr[(\hat A-\hat B)^{\dagger}(\hat A-\hat B)]^{1/2}$.

Let us then write the output states
as
\bea
\hat\rho_1&=&p\proj{g_1}+(1-p)\proj{g_2},
\nonumber \\
\hat\rho_2&=&q\proj{g_2}+(1-q)\proj{g_1},
\label{output}
\eea
with the output probabilities $(p, q)\in [0, 1]$.
The above condition Eq. (\ref{au}) implies
a complicated set of constraints on the parameters
describing generical mixed input and output states,
and on the probability distributions $p, q$,
but it can be explicitly calculated within a nice geometrical framework.

In particular, in the most general model of mixed `initial' and 
`template' states defined by an arbitrary vector in the Bloch sphere,
$\hat f_i\equiv [\hat I+ \vec{f_i}\cdot \hat \sigma]/2$
and $\hat g_i \equiv [\hat I+ \vec{g_i}\cdot \hat \sigma]/2$,
respectively, Alberti and Uhlmann's condition can be rewritten as

\bea
h(\hat p, \hat q; \vec{f_i}, \vec{g_i}; t)&\equiv &
h^B-|h^B|
\nonumber \\
&-&R(h^A-|h^A|)\geq 0~~~;~~~\mbox{$\forall t\in{\cal R}^+$}
\label{hm}
\eea
where, using the new coordinates $\hat p\equiv p-1/2$, $\hat q\equiv
q-1/2$ ($(\hat p, \hat q)\in [-1/2, 1/2]$) to simplify the notation,
we have introduced the parabolic functions of $t$ as

\bea
h^A(X; t)&\equiv &X-2(2+X)t+Xt^2
\nonumber \\
h^B(\hat p, \hat q; Y_0; t)&\equiv &(Y_0-4\hat p^2)-2(Y_0+4\hat p\hat
q)t 
\nonumber \\
&+&(Y_0-4\hat q^2)t^2,
\label{hahb}
\eea
and the parameters

\bea
R&\equiv &{f^2\sin^2\phi\over g^2\sin^2\theta}~\geq 0, 
\nonumber \\
X&\equiv & {1-f^2\over f^2\sin^2\phi}~\geq 0, 
\nonumber \\
Y_0&\equiv &1+ {1-g^2\over g^2\sin^2\theta}~\geq 1,
\label{parameter}
\eea
with $2\sin^2\theta\equiv 1-\vec{g_1}\cdot\vec{g_2}/g^2$,  
$~2\sin^2\phi\equiv 1-\vec{f_1}\cdot\vec{f_2}/f^2$ and $\phi, \theta 
\in [0, \pi]$.  

Now let us turn to the analysis of condition \rf{hm}.
This can be seen to reduce to the following constraints 
\bea 
\Delta 
t_+&\equiv &t_+^A-t_+^B\geq 0, \nonumber \\
\Delta t_-&\equiv &t_-^B-t_-^A\geq 0,
\label{deltat}
\eea
where $t^A_{\pm}$ and $t^B_{\pm}$ are the zeros of $h^A$ and $h^B$,
respectively, and

\bea 
H(\hat p,\hat q, R, X, Y_0; t)&=&(Y_{0X}-4\hat 
p^2)-2[Y_{2X}+4\hat p\hat q]t \nonumber \\
&+& (Y_{0X}-4\hat q^2)t^2\geq 0~~~;
\nonumber \\
&~&~~\mbox{for $t_-^B\leq t\leq t_+^B$},
\label{deltah}
\eea
where, for ease of presentation, we have defined $Y_{nX}\equiv Y_0-(n+X)R$.
After some algebra and the analysis of a few geometrical constraints in the
parameter space $(p, q)$, 
one finally obtains that the Alberti-Uhlmann condition can be satisfied
in certain geometrically simple $(p, q)$ parameter regions, classified 
according to the values of $R, X$ and $Y_0$ (see Appendix).

\section{Repeater in Lossy Quantum Channel}

In the model for the repeater in a quantum lossy channel, 
the input states are pure, and
the Alberti and Uhlmann condition can be greatly simplified as
the well known fidelity criterion \cite{uhlmann}
\beq
F(\hat f_1, \hat f_2)\leq F(\hat \rho_1, \hat \rho_2).
\label{fid}
\eeq

Given the output states \rf{output},
it is easy to evaluate the fidelities so that
the CPTP mapping existence condition \rf{fid} can be
explicitly rewritten as
\bea
pq+(1-p)(1-q)-R
\le
2 \sqrt{p(1-p)q(1-q)},
\label{cpcond}
\eea
where we have introduced the parameters
\beq
R\equiv
\frac{1-\kappa^2}{1-K^2}<1,
\quad
\kappa\equiv\langle\alpha\vert-\alpha\rangle,
\quad
K\equiv\langle\beta\vert-\beta\rangle.
\label{ab}
\eeq

\begin{figure}
\begin{center}
\includegraphics[width=0.45\textwidth, height=6cm]{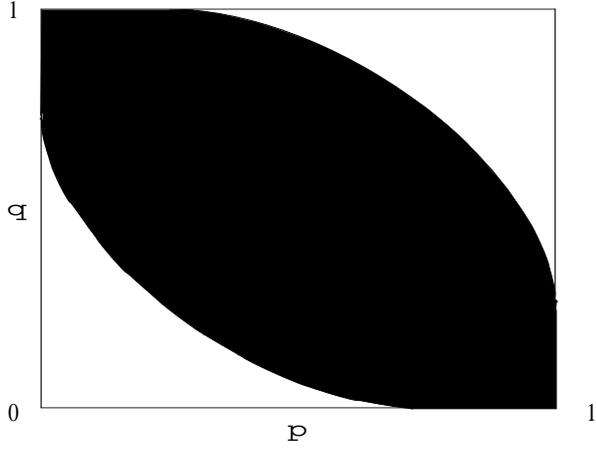}
\end{center}
\caption{\label{fig:allowed region}
The allowed $(p, q)$ region (shaded area) for the existence of a CPTP mapping
between two input pure states and two output mixed states ($R=0.25$).}
\end{figure}

The inequality \rf{cpcond} is trivially satisfied
when its l.h.s. is negative definite,
i.e. when
\begin{eqnarray}
q
&\ge&
{1\over2}\left [{2R-1 \over 2p-1}+1\right ]
\quad
(0<p<{1\over2}), \\
q
&\le&
{1\over2}\left [{2R-1 \over 2p-1}+1\right ]
\quad
({1\over2}<p<1).
\label{hyperbole}
\end{eqnarray}
Otherwise,
\beq
\Delta(p, q)\equiv (p+q+R-1)^2-4Rpq
\le
0,
\label{hypell}
\eeq
should hold.
Collecting these two cases together,
we finally conclude that
the CPTP mapping existence condition \rf{cpcond}
is satisfied for the range of parameters $(p, q)$
contained within the shaded area shown in
Fig. \ref{fig:allowed region}.
The upper boundary is specified by
\beq
\begin{array}{ll}
q&=1 \qquad\qquad (0\le p\le R), \\
q&=1-R-(1-2R)p+2\sqrt{R(1-R)p(1-p)} {} \\
{} & \qquad\qquad\qquad (R\le p\le1), \\
0&\le q\le R \qquad (\mathrm{at}\quad p=1),
\label{upper boundary}
\end{array}
\eeq
while the lower boundary is given by
\beq
\begin{array}{ll}
1&-R\le q\le 1 \qquad (\mathrm{at}\quad p=0), \\
q&=1-R-(1-2R)p-2\sqrt{R(1-R)p(1-p)} {} \\
{} & \qquad\qquad\qquad (0\le p\le 1-R), \\
q&=0 \qquad\qquad (1-R\le p\le1).
\label{lower boundary}
\end{array}
\eeq

\subsection{Optimal Repeater}\label{Optimal repeater}

Now we apply the above results to derive the optimal repeater
for the second channel
which is assumed to be a simple lossy channel described by
\beq
\hat{\cal L} (\proj{\pm\gamma})=\proj{\pm\eta\gamma},
\label{lossy channel}
\eeq
for any coherent state $\ket{\gamma}$ and $0<\eta<1$.
We consider two kinds of measures of the transmission performance
through the lossy channel, i.e.
the average bit error rate $P_{\mathrm{e}}$
and
the Holevo capacity $\chi({\cal E})$
for the output ensemble from the channel,
${\cal E}=\{\hat\rho_1',\hat\rho_2';1-\xi,\xi\}$,
where
$\hat\rho_i'\equiv\hat{\cal L}(\hat\rho_i)$ and
\begin{eqnarray}
\hat\rho_1
&\equiv&
p\proj{\beta}+(1-p)\proj{-\beta},
\nonumber \\
\hat\rho_2
&\equiv&
q\proj{-\beta}+(1-q)\proj{\beta},
\label{rho}
\end{eqnarray}
$1-\xi$ and $\xi$ being the {\it a priori} probabilities
for $\hat\rho_1'$ and $\hat\rho_2'$, respectively,
as well as for $\ket{\alpha}$ and $\ket{-\alpha}$.

We first consider minimizing the average error probability
$P_{\mathrm{e}}$ with respect to a POVM $\{\hat\Pi_1, \hat\Pi_2\}$
\bea
P_{\mathrm{e}}^\mathrm{min}
&\equiv&
\min_{\{\hat\Pi_1, \hat\Pi_2\}}
\left[
(1-\xi) \tr (\hat \Pi_1\hat\rho_1')
+
\xi\tr(\hat \Pi_2\hat\rho_2')
\right]
\nonumber \\
&=&
1-\xi +
\min_{\hat \Pi_{1}} \left[\tr(\hat\Pi_1\hat\Lambda)\right ],
\label{helstrom}
\eea
where $\hat\Lambda\equiv \xi \hat\rho_2'-(1-\xi)\hat\rho_1'$
and we have used the property $\hat \Pi_1+\hat \Pi_2=\hat I$.
The minimum error is then found by taking 
$\hat\Pi_1=\ket{\lambda_-}\bra{\lambda_-}$,
where $\ket{\lambda_-}$ is
the negative eigenvalue eigenstate of the operator $\hat\Lambda$.
We then have
\begin{eqnarray}
P_{\mathrm{e}}^\mathrm{min}(p, q)
&=&{1\over2}
\Biggl\{
1-
\biggl[
(2\xi-1)^2K'^2
\nonumber\\
&+&[2\xi q+2(1-\xi)p-1]^2(1-K'^2)
\biggr]^{1\over2}
\Biggr\},
\eea
with $K'\equiv \langle \eta\beta\vert -\eta\beta\rangle$.
So clearly we are to find the optimal repeater maximizing
the quantity 
\beq
S(p, q)\equiv\xi q+(1-\xi)p.
\eeq
Since the latter is an increasing function in both $p$ and $q$,
it can be maximized under the CPTP map existence constraints
by use of the standard Lagrange multiplier method,
i.e. by solving the following set of equations
\beq
\nabla S(p, q)
=\lambda \nabla [\Delta(p, q)]
~~~;~~~\Delta(p, q)=0,
\label{lagrangepure}
\eeq
where $\lambda$ is a constant.
Solving Eqs.  \rf{lagrangepure} with the aid of Eq. \rf{hypell} and
Fig. 1, it is readily shown that the optimal bit error rate
is obtained for:
\beq
p_\mathrm{opt}={1\over 2}\left [1+{c_-\over \sqrt{c}}\right ]~~~;~~~
q_\mathrm{opt}={1\over 2}\left [1+{c_+\over \sqrt{c}}\right ],
\label{optpairpure1}
\eeq
for $0<R<1$ and $0<\xi<1$, where

\bea
c_{\pm}&\equiv &R\pm (2\xi-1)(1-R),
\nonumber \\
c&\equiv & 1-4\xi(1-\xi)(1-R).
\label{parameters}
\eea
Furthermore, for the optimal pair \rf{optpairpure1} we have
\beq
S_\mathrm{opt}=C+D[\sqrt{c}-1]/2.
\eeq
Note that in the particular case of equiprobably distributed inputs, 
i.e.
when $\xi=1/2$, we have that $c_+=c_-=c=R$ and then the optimal
point for $0<R<1$ explicitly reads
$p_\mathrm{opt}=q_\mathrm{opt}=(1+\sqrt{R})/2$.

With $(p, q)$
evaluated as the optimal pair \rf{optpairpure1} we get
\beq
P_{\mathrm{e, CPTP}}^{\mathrm{min}}
={1\over 2}\{1-\sqrt{1-4\xi(1-\xi)[1-(1-K'^2)R]}\}.
\label{errcptp}
\eeq
We compare this with the average bit error rate 
in the case of no action by the repeater, 
i.e. with final states given by $\proj{\pm\eta\alpha}$,
which is expressed by 
\beq
P_{\mathrm{e, NO ACT}}\equiv [1-\sqrt{1-4\xi(1-\xi)\kappa'^2}]/2.
\label{err_noact}
\eeq
where $\kappa'\equiv \langle \eta\alpha\vert -\eta\alpha\rangle$.
As it can be simply proved and directly seen from Fig.  2,
the optimal error probability $P_{\mathrm{e, CPTP}}^{\mathrm{min}}$ 
is always smaller than $P_{\mathrm{e, NO ACT}}$
for any choice of initial probability distributions
$\xi$, $0<\eta<1$ and
$|\beta|>|\alpha|$.
That is, the intermediate action of the repeater with optimal CPTP mapping
on the initial states reduces the final error probability of
detecting the original states.

\begin{figure}
\begin{center}
\includegraphics[width=0.45\textwidth, height=6cm]{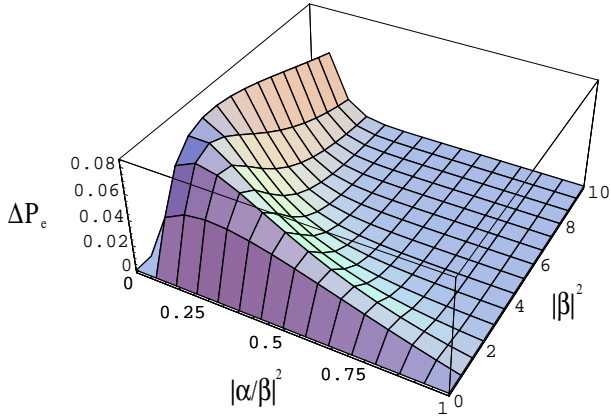}
\end{center}
\caption{\label{Fig. 2}
The difference in the error probabilities for Bob, 
$\Delta P_{\mathrm{e}}
\equiv 
P_{\mathrm{e, NO ACT}}-P_{\mathrm{e, CPTP}}^{\mathrm{min}}$, 
as a function of $|\alpha/\beta|^2$ and $|\beta|^2$
in the case $\eta=1/\sqrt{2}$, $\xi=1/2$.}
\end{figure}

Now we turn our attention to the problem of maximizing the
Holevo capacity
\beq
\chi({\cal E})
\equiv S(\hat\rho')-\sum_k \xi_k S(\hat\rho_k')
=\sum_k \xi_k D(\hat\rho_k' \vert\vert \hat\rho'),
\eeq
where $\hat \rho'=\sum_k \xi_k \hat\rho_k'$,
$S(\hat\rho')$ is the von Neumann entropy,
and
$D(\hat\rho_k' \vert\vert \hat\rho')$ is the relative entropy.
First notice that
$\chi({\cal E})$ is maximized at the extreme points of
the convex set $(p,q)$ of the region allowed by the Alberti-Uhlmann
condition, 
because $\chi({\cal E})$ is a downward convex function with respect 
to the pair $(p,q)$. 
In fact,
let
$(p_\mathrm{E},q_\mathrm{E})$
and
$(p_\mathrm{A},q_\mathrm{A})$
be extreme and interior points, respectively.
Define the corresponding ensembles as
${\cal E}^\mathrm{E}=\{\hat\rho_k^\mathrm{E};\xi_k\}$
and
${\cal E}^\mathrm{A}=\{\hat\rho_k^\mathrm{A};\xi_k\}$.
Then for another interior point,
\beq
\hat\rho_k^\mathrm{B}
=(1-\zeta)\hat\rho_k^\mathrm{E}
+\zeta\hat\rho_k^\mathrm{A}
\eeq
(where $0<\zeta<1$), we have
\beq
\chi({\cal E}^\mathrm{B})
\le
(1-\zeta)\chi({\cal E}^\mathrm{E})+\zeta\chi({\cal E}^\mathrm{A}),
\eeq
due to the joint convexity of the relative entropy.

The problem of maximization of the Holevo capacity 
$\chi_{\mathrm{CPTP}}$ along
the (elliptic) boundary of the CPTP allowed region in the $(p, q)$
parameter space for general initial probability distributions $\xi$
is still quite cumbersome but can be solved numerically.
For the sake of clarity we explicitly show here a practical case of
equiprobably distributed inputs, $\xi=1/2$ (maximum amount of
information encoded in the inputs).
It is quite easy to check that in this case the channel capacity
is zero along the line $q=1-p$ and symmetric with respect to the
lines $q=p$ and $q=1-p$, and monotonically increasing towards the 
points $(1, 1)$ and $(0, 0)$.
In particular, its maximum is achieved at the optimal point 
$p_\mathrm{opt}=q_\mathrm{opt}=
(1+\sqrt{R})/2$ on the boundary of the allowed region.
Its behaviour as a function of $\kappa$ is shown in Fig. 3, where
it is also compared with the channel capacity 
\beq
\chi_{\mathrm{NO ACT}}
\equiv -(
\lambda_{\kappa +}\log \lambda_{\kappa +}+\lambda_{\kappa -}\log 
\lambda_{\kappa -})
\eeq
(with $\lambda_{\kappa \pm}\equiv (1\pm \kappa )/2$)
for the case of no action by the repeater and the Holevo bound for the
input states, 
i.e. 
\beq
\chi_{\mathrm{INPUT}}
\equiv -(
\lambda_{\kappa' +}\log \lambda_{\kappa' +}
+\lambda_{\kappa' -}\log \lambda_{\kappa' -})
\eeq 
(with $\lambda_{\kappa' \pm}\equiv (1\pm \kappa')/2$).
\begin{figure}
\begin{center}
\includegraphics[width=0.45\textwidth, height=6cm]{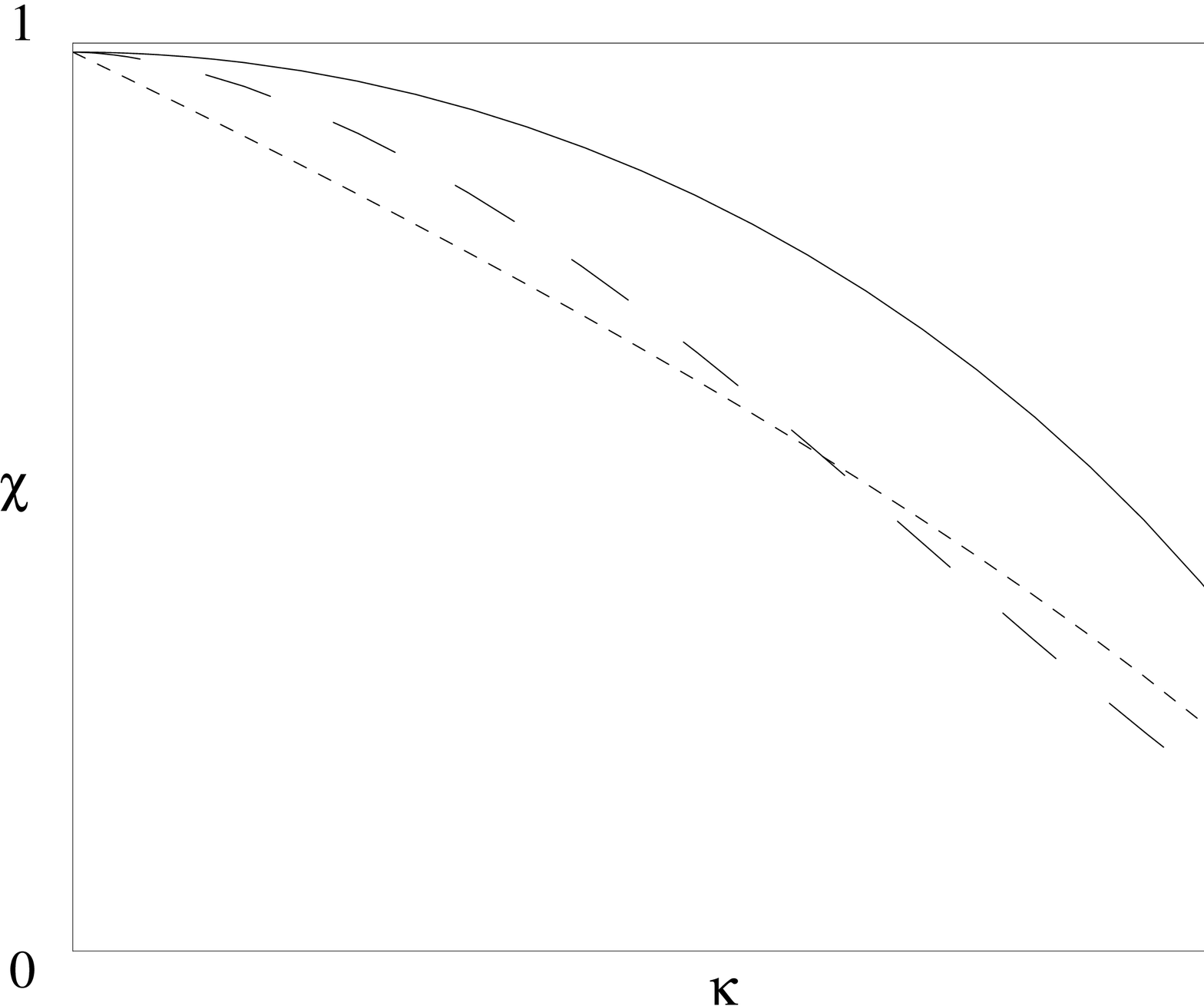}
\end{center}
\caption{\label{fig:Holevo bounds at xi=1/2}
The Holevo capacities $\chi_{\mathrm{CPTP}}$ (dashed line), 
$\chi_{\mathrm{NO ACT}}$ (dotted line) and 
$\chi_{\mathrm{INPUT}}$ (continuous line)
as a function of the inputs overlap $\kappa$ 
for the equiprobable inputs case ($\xi=1/2$), 
$|\beta/\alpha|=2$ and $\eta=1/\sqrt{2}$.}
\end{figure}
As one can see, there are both parameter ($\beta/\alpha, \eta$) regions 
where $\chi_{\mathrm {CPTP}}>\chi_{\mathrm {NO ACT}}$ 
and $\chi_{\mathrm {CPTP}}<\chi_{\mathrm {NO ACT}}$.
In particular, defining $\kappa_0$ as the
intercept point between the curves $\chi_{\mathrm{CPTP}}$ and 
$\chi_{\mathrm{NO ACT}}$ (i.e. such that $\chi_{\mathrm{CPTP}}(\kappa_0)
\equiv \chi_{\mathrm{NO ACT}}(\kappa_0)$), for $0<\kappa<\kappa_0<1$
the accessible information
is bigger when amplifying the signals at the repeater, while for
$\kappa_0<\kappa<1$ the best performance is obtained without amplification. 
This behavior can be explained as follows:
for small $\kappa$ the inputs tend to be more orthogonal 
and the quantum repeater helps;
on the other hand, for larger $\kappa$, the inputs tend to overlap 
and there is no gain in using the quantum repeater. 
Furthermore, one can easily check that, as $\eta$ decreases (the channel
becomes more lossy), although the absolute channel capacity performance
decreases, the range of $\kappa$ for which $\chi_{\mathrm{CPTP}}>
\chi_{\mathrm{NO ACT}}$
also becomes larger ($\kappa_0$ increases): for very noisy channels
the amplification by the repeater is essential even for the case when
the inputs are almost completely overlapping. 

The different behavior measured by the minimum bit error rate and 
the Holevo capacity may be also interpreted as follows: 
the Helstrom bound specifies the performance of 
a single shot measurement on each signal state,
while the Holevo capacity is a measure for the coding 
by a large scale collective measurement 
where the coherence involved in sequences of signal states 
must be fully used to extract as much information as possible.
So, preparing mixed state signals at the repeater could spoil 
in some cases 
the coherence involved in sequences of pure state signals 
$|\pm \eta\alpha\rangle$, 
leading to the reduction of the Holevo capacity. 

For the near future optical communications 
based on classical coding, 
the bit error rate is of greater interest, 
and the optimal repeater derived here will be useful. 
When the template states 
$\{\vert\beta\rangle, \vert-\beta\rangle\}$ 
can be prepared with enough power such as $K\sim 0$, 
then the optimal repeating strategy is simply realized 
by the intercept-resend (IR) strategy. 
That is, 
we first discriminate $\{ \vert\beta\rangle, \vert-\beta\rangle \}$ 
by the minimum error measurement, 
and then assign an appropriate template state based on
the measurement results. 
In the case of $\xi=1/2$, the repeating states are specified by 
Eq. (\ref{rho}) with the parameters 
\beq
p=q={1\over2}\left( 1+\sqrt{1-\kappa^2} \right), 
\eeq
and the final bit error rate is 
\beq
P_{\mathrm{e, IR}}^{\mathrm{min}}
={1\over 2}\{1-\sqrt{(1-\kappa^2)(1-K'^2)]}\}.
\eeq

\begin{figure}
\begin{center}
\includegraphics[width=0.45\textwidth, height=6cm]{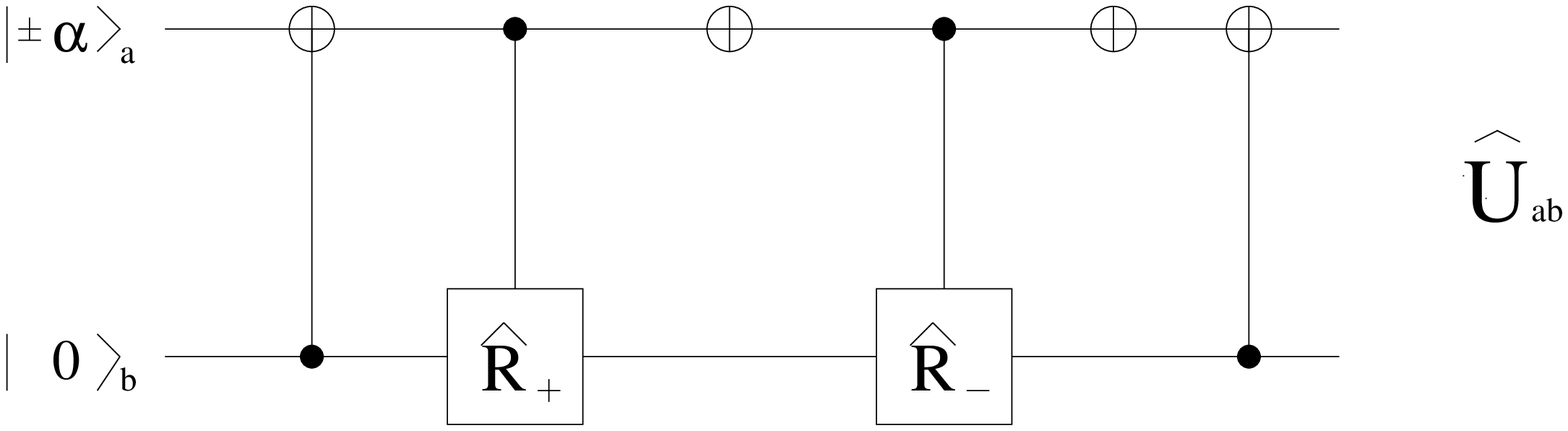}
\end{center}
\caption{\label{fig:CPTP network}
The network which realizes the optimal CPTP mapping for the repeater,
with ${\hat U}_{ab}\equiv \exp\{ \ket{0}_a\bra{1}[\theta_-\ket{0}_b\bra{1} 
+ \theta_+\ket{1}_b\bra{0}]-\mathrm{h.c.}\}$.}
\end{figure}

When, on the other hand, 
the non-orthogonality of the template states 
should be taken into account, 
we have to consider quantum processes which do not include any 
intermediate measurement process. 
One possible implementation is given by the quantum network 
shown in Fig. 4. 
The computational basis is made up of the so called
even and odd coherent states, 
\bea
|0\rangle&\equiv& {1\over \sqrt{2(1+\kappa)}}
(|\alpha\rangle +|-\alpha\rangle),
\nonumber \\
|1\rangle&\equiv& {1\over \sqrt{2(1-\kappa)}}
(|\alpha\rangle -|-\alpha\rangle).
\label{evenodd}
\eea
The controlled rotations are defined by 
\bea
\hat R(\theta_{\pm})\equiv
\left(\begin{array}{cc}
\cos\theta_{\pm}&-\sin\theta_{\pm}\\
\sin\theta_{\pm}&\cos\theta_{\pm}
\end{array}\right ),
\eea
where $\theta_{\pm}\equiv (\arcsin K/\kappa\pm\pi/2)/2$. 
The ancilla qubit is initialized   
in the even coherent state $|0\rangle_b$, 
where the subscript $b$ refers to a
particular mode of the coherent template states.
The repeating states are simply obtained at the output port 
in mode b 
by tracing out the states in mode $a$. 
Unfortunately, this type of quantum circuit still requires 
hypothetical non-linear processes to generate 
the even and odd coherent states as well as the cross Kerr effect 
between mode a and b \cite{sasaki,Cochrane99}.

\section{State-dependent Quantum Cloning}

Another interesting application of the CPTP mapping results is in 
state-dependent cloning.
As it is well known, an arbitrary unknown quantum state cannot be 
cloned \cite{wootterszurek}.
It is possible, however, to produce imperfect copies of quantum states,
both deterministically (when the cloning machine can only perform unitary 
operations) and probabilistically (where via postselection measurements in an 
ancillary space, faithful copies of the input are obtained with non zero 
success probability).
Several results on quantum cloning are already known by now 
(for a selected, though not exhaustive, bibliography, see, e.g.,
Refs. \cite{fanmatsumotowadati}).
 
In this section we will exploit the geometric results concerning the
existence of a CPTP map between 2-d quantum systems section to describe
an $N\rightarrow M$ (anti) cloning state-dependent machine.  In 
particular, we assume that the input states are pure and given as an 
$N$-fold tensor product $\ket{f_i}^{\otimes N}$, while the `templates' 
$\hat g_i$  are pure ($g=1$) and $M$-copies 
clones ($M\geq N\geq 1$) of the input states $\ket{f_i}$, i.e.  

\beq 
\ket{f_i}\rightarrow \ket{\tilde f_i}\equiv \ket{f_i}^{\otimes 
N}~~~;~~~ \ket{g_i}\equiv\ket{f_i}^{\otimes M}.
\label{gcloning}
\eeq

We then restrict our analysis to the special case in which 
we assumed that we are only able to construct outputs which 
are classical mixtures of these templates, that is the outputs
are given again by Eq. \rf{output}.
More general cloner models (including the state-dependent copiers
which unitarily map pure initial states to a pure state superposition
of clones as in Refs. \cite{bruss1,chefles2}) will be considered elsewhere.
In our ansatz, then, it is straightforward to see, by using the 
Bloch sphere parametrization for $\hat g_i$
and noting that the states $\{\ket{g_i}\}$ (as well as the states
$\{\ket{f_i}\}$) span a 2-d Hilbert space,
that the overlaps must be
\beq
|\left\langle \tilde f_1|\tilde f_2\right\rangle| 
=\cos^N\phi~~;~~
|\left\langle g_1|g_2\right\rangle |= \cos\theta = \cos^M\phi.
\label{angle} 
\eeq 
The case of pure $\ket{g_i}$ can be also immediately handled 
within the framework discussed in the previous section provided that 
we take $Y_0=1$ (see Eq.  \rf{parameters}).  Therefore, for the 
parameter $R$ of Eq.  \rf{parameter}, we obtain 
\beq 
R={1-\cos^{2N}\phi\over 
1-\cos^{2M}\phi},
\label{apure}
\eeq
with
$R\in [N/M, 1]$.
In order to evaluate the efficiency of the cloning machine, we can
now either choose as the figure of merit the `global' fidelity (see, 
e.g., Refs.  \cite{bruss1,chefles2})
\beq
\bar F_G\equiv (1-\xi)\left\langle \tilde f_1|\hat \rho_1| \tilde 
f_1\right\rangle +\xi \left\langle \tilde f_2|\hat \rho_2| \tilde 
f_2\right \rangle,
\label{globalscore}
\eeq
which can be easily seen to correspond (taking $g=Y_0=1$, and $\theta$
and $R$ as defined in Eqs.  \rf{angle}-\rf{apure}) to
\beq
\bar F_G= Z^M+(1-Z^M)[(1-\xi)p+\xi q],
\eeq
with $Z\equiv \cos^2\phi$, and
then essentially the same as the score $S(p, q)$ of the previous section,
with the same maximum at the optimal points $(p_{opt}, q_{opt})$ of
Eq.  \rf{optpairpure1}, finally giving (note that for cloning, $R<1$, see Eq.  
\rf{apure} and Fig.  2a, and the condition $Y_0=1$ also implies that 
$\xi_0=0$): \bea \bar F_{G, opt}(Z; \xi,& N&, M)=1-{(1-Z^M)\over 2} 
\nonumber \\
&\cdot &
\biggl [1-\sqrt{1-4\xi(1-\xi){(Z^N-Z^M)\over (1-Z^{M})}}\biggr ].
\label{globalscore2}
\eea
Otherwise, we could choose the `local' fidelity (see, e.g., Refs.
\cite{bruss1,chefles2})

\beq
\bar F_L \equiv (1-\xi)F_1(\hat f_1, \hat f_1^{out}) +\xi 
F_2(\hat f_2, \hat f_2^{out}),
\label{clonescore}
\eeq 
where $F_i(\hat f_i, \hat f_i^{out})$ is the fidelity between the 
reduced density operator for one single copy of the initial state 
(i.e., $\hat f_i$) and the reduced density operator for one single 
copy of the final state (i.e., $\hat f_i^{out}$, obtained tracing out 
any $M-1$ qubits from $\hat \rho_i$, and which is independent of the 
choice of the remaining copy).  Since the 
output reduced density operators (cf.  Eq.  \rf{output}) are given by
\bea \hat f_1^{out}&=& p\hat f_1 +(1-p)\hat f_2, \nonumber \\
\hat f_2^{out}&=& q\hat f_2 +(1-q)\hat f_1,
\label{reducedop}
\eea
a short calculation shows that 
\beq
\bar F_L= Z+(1-Z)[(1-\xi)p+\xi q],
\eeq 
which is again optimized by the parameters of Eq.  \rf{optpairpure1}
and finally reads
\bea
&~&(1-Z^M)(1- \bar F_{L, opt}(Z; \xi, N, M))=
\nonumber \\
&~&(1-Z)(1-\bar F_{G, opt}(Z; \xi, N, M)).
\label{clonescore2}
\eea
Since the `local' and `global' fidelities are linearly 
correlated, it is enough in the following to study the behaviour of 
one of them, e.g.  $\bar F_L$.  First of all, cloning is not allowed 
for the set of parameters $(p, q)$ outside the shaded region of Fig.  
2a.  Then, considered as a function of $\xi$, $\bar F_{L, opt}$ is 
further maximized (as expected) for the trivial choices $\xi=0$ or 
$\xi=1$ (only one `input' state), for which $\bar F_{L, opt}=1$.  It 
is also easy to see that the optimal $\bar F_{L, opt}(\xi)$ is bounded 
below by $\bar F_{L, opt}(\xi=1/2)$, i.e.  for the choice of 
equiprobabilistically distributed input states $\{\ket{f_i}\}$.  This 
case is important because for $\xi=1/2$ the maximum amount of 
information is encoded in the input states.  It is easily seen that 
this fidelity is an increasing function of $N$ and a decreasing 
function of $M$.  As a function of $Z$ at fixed $N, M$ it decreases 
from the maximum $\bar F_{L, opt}(Z; 1/2, N, M)=1$ at $\phi=0$ (the 
case for maximally indistinguishable initial states) until it reaches 
a minimum around $\phi_{min} \geq \pi/4$ (for $N=1$ and $M=2$, at 
which $\bar F_{L, opt}\simeq 0.95$) and then again increases towards 
$\bar F_{L, opt}(Z; 1/2, N, M)=1$ at $\phi=\pi/2$ (the case for 
orthogonal, classical inputs).  In the asymptotic case of 
$M\rightarrow \infty$ the `local' fidelity has a similar shape, with 
the minimum (for $N=1$) $\bar F_{L, opt} =25/27\simeq 0.92$ at 
$\phi_{min}=\arccos \sqrt{5/9}\leq \pi/4$.  The optimal `local' 
average fidelity $\bar F_{L, opt}(Z; 1/2, N, M)$ is plotted as a 
function of $Z$ for $N=1$ and $M=2, \infty$ in Fig.  4.  Also note 
that, in the asymptotic limit of $M\rightarrow \infty$, the `global' 
fidelity reaches the Helstrom bound 
\beq
2\bar F_{\mathrm{Helstrom}} 
\equiv 1+\left [1-4\xi(1-\xi)\left\langle \tilde f_1|\tilde f_2
\right\rangle \right ]^{1/2},
\eeq
which is the maximum probability to 
distinguish the two states $\ket{\tilde f_1}$ and $\ket{\tilde f_2}$.  
Quantum cloners with state dependent fidelity were already considered 
in the literature, see, e.g., Refs.  
\cite{bruss1,chefles2,buzekhillery1}.  One of their most important 
practical use is for eavesdropping strategies in some quantum 
cryptographic system.  As Fig. 4 shows, our local and global 
fidelities for $\xi=1/2$ are smaller than, respectively, the optimal 
eavesdropping strategy fidelity described in Ref.  \cite{bruss1} and 
the global one of Ref.  \cite{chefles2}.
As we have already stressed,
this is just a consequence of the peculiarity of our output states,
which are a classical mixture of the perfect clones $\ket{f_i}^{\otimes M}$,
while in Refs.  \cite{bruss1,chefles2} the optimization is over
a unitary transformation between arbitrary initial and final pure states.
The evident advantage of our optimal CPTP mapping method in a general 
cloning machine relies in not having to deal with all the inequalities 
which derive 
from the constraints on the unitarity of transformations over extended 
Hilbert spaces with ancilla qubits, as we just have to maximize the
chosen figure of merit along a certain curve specifying the boundary of
the allowed CPTP mappings between the initial and the output (mixed) states.

\begin{figure}
\begin{center}
\includegraphics[width=0.45\textwidth, height=6cm]{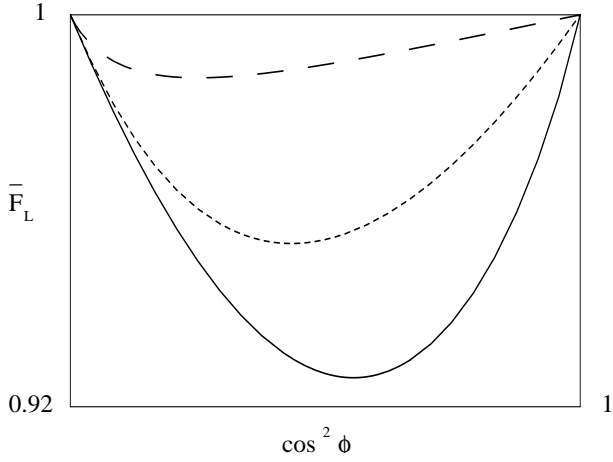}
\end{center}
\caption{\label{Fig. 4}
The optimal average score $\bar F_{opt}(Z; 0.5, 1, M)$ for the parameters:
a) $M=2$ (dotted line); b) $M=\infty$ (continuous line),
and the optimal local eavesdropping strategy fidelity $F_{l, 3}$
in Eq. (51) of Ref. \cite{bruss1} for $N=1$ and $M=2$ (dashed line).}
\end{figure}

The importance and relation of different `quality' measures for cloning other
than fidelity, and for instance the realization that generally copiers
quantum optimized with respect to fidelity are not optimal with respect
to information transfer measures, and viceversa, was stressed, e.g.,
in Ref. \cite{deuar,footnote}.
In particular, another measure of the quality of the performance 
of our copier can be given in terms of the Holevo bound on the copied 
information for the reduced density outputs \rf{reducedop}, i.e. (for the
optimal point given by Eq. \rf{optpairpure1})
\bea
I_H(Z; \xi, N, M)&\equiv &S(\sum_i p_i \hat f_i^{out})
-\sum_i p_i S(\hat f_i^{out})
\nonumber \\
&=&\sum_{\alpha=\pm ; i=1,2,3}P_i~\lambda^{i}_{\alpha}\log \lambda^{i}_{\alpha}
\label{holevobound}
\eea
where 
\bea
2 \lambda_{\pm}^{1}&\equiv &1\pm \{[c_-^2+4\xi^2R(1-R)Z]/c\}^{1/2},
\nonumber \\
2 \lambda_{\pm}^{2}&\equiv &1\pm \{[c_+^2+4(1-\xi)^2R(1-R)Z]/c\}^{1/2},
\nonumber \\
2 \lambda_{\pm}^{3}&\equiv &1\pm \{[(1-2\xi)^2+4\xi(1-\xi)RZ]/c\}^{1/2},
\eea 
$P_1=1-\xi, P_2=\xi, P_3=-1$ and $c_{\pm}, c$ and $R$ are given, 
respectively, by Eqs. \rf{parameters} and \rf{apure}.
This should be compared with the maximum information extractable from the
original states, given by
\beq
I_H^{in}(Z; \xi)\equiv S(\sum_i p_i \hat f_i) 
=-\sum_{\alpha=\pm}\lambda^{in}_{\alpha}\log \lambda^{in}_{\alpha}, 
\label{ihinput}
\eeq
with 
\beq
2 \lambda_{\pm}^{in}\equiv 1\pm [(1-2\xi)^2+4\xi(1-\xi)Z]^{1/2}.
\eeq
These figures of merit are shown in Fig. 5 for $\xi=1/2$, $N=1$
and $M=2, \infty$, and compared with the Holevo bound of the
Wooters and Zurek model \cite{wootterszurek} (which, in this sense,
is nearly optimal as it allows to extract as much information from 
the copies as from the originals \cite{deuar}).

\begin{figure}
\begin{center}
\includegraphics[width=0.45\textwidth, height=6cm]{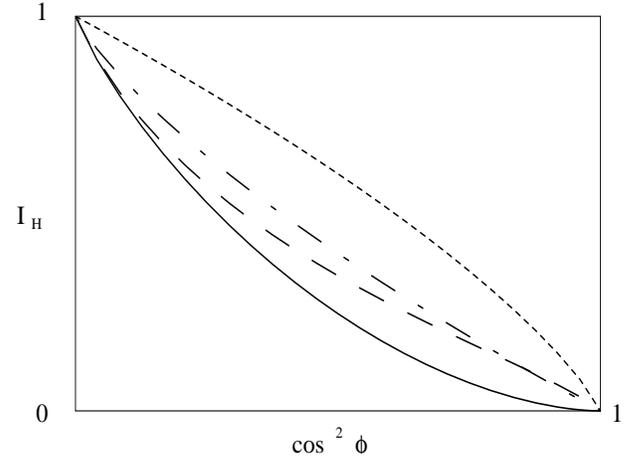}
\end{center}
\caption{\label{Fig. 5}
The Holevo bound on the copied information $I_H(Z; 0.5, 1, M)$, 
for the parameters $M=2$ (dashed line) and $M=\infty$ (continuous line),
compared to the Holevo bound for the Wooters Zurek (WZ) cloner (dot-dashed line)
and the maximal information extractable from the input states,
$I_H^{in}(Z; 0.5)$ (dotted line).} 
\end{figure}

Completely similar considerations can be extended to the case in which the
input and the template states are, respectively, the coherent states 
$\ket{\pm \alpha}$ and $\ket{\pm \beta}$, just by replacing in the previous 
formulas for $Z\rightarrow Z_{coh}\equiv|\left \langle\alpha|
-\alpha\right\rangle|^2=Exp[-4|\alpha|^2]$.
Furthermore, with the same methods we can also consider a special type of 
copier called $N\rightarrow K+L$ (with $K+L\geq N$) `anti-cloning' 
machine \cite{songhardy}.
In this ansatz, a set of unknown `input' states $\{\ket{f_i}^{\otimes N}\}$
is transformed into the tensor
product of $K$ copies of the input $\ket{f_i}$ times $L$ copies of a 
state $\ket{-f_i}\equiv\bar \beta_i\ket{0}-\bar \alpha_i\ket{1}$
which has opposite spin direction with respect to the input one.
This type of cloning is physically interesting for a number of information
theoretic reasons (see, e.g., Refs. \cite{gisinpopescu}). 

The pure `templates' are thus chosen as
$\ket{g_i}\equiv \ket{f_i}^{\otimes K}\ket{-f_i}^{\otimes L}$,
such that now
$|\left\langle g_1|g_2\right \rangle| 
=\cos^{K+L}\phi$  and 
$R=[1-\cos^{2N}\phi]/[1-\cos^{2(K+L)}\phi]$,
with $R\in [N/(K+L), 1]$.
The analysis of the optimal efficiency of the anti-cloning machine
then follows similar lines to those of the previous cloning machine case, 
just provided that one makes the substitution $M\rightarrow K+L$.

\section{Discussion}

We have considered the constraints on the existence of CPTP mappings
between two arbitrary initial pure states and two arbitrary final mixed
states using Uhlmann's theorem \cite{uhlmann} and interpreting them
within a simple geometrical picture.
Exploiting these results, we then studied the model of a quantum communication 
channel where a set of coherent states are sent by Alice, eventually transformed
by an intermediate repeater who can perform an optimal CPTP 
mapping and, after going through a lossy channel ${\cal L}$, 
are finally received by Bob 
with a certain error probability.  We have shown that when the 
intermediate repeater performs the optimally CPTP mapping, the final error 
probability is always smaller than in the case when no action is taken 
at the intermediate stage.  In other words, we can have a gain when 
the optimal mapping strategy is applied to repeat or amplify the input 
signals in the channel.  This is a new and intriguing result for quantum 
communication, showing the potential relevance of the optimal CPTP mapping 
strategy.

Furthermore, the optimal CPTP mapping constraints have been used to 
analyze state-dependent optimal cloners where the output is a classical mixture 
of exact copies of the initial inputs, and the `local' and `global' fidelity
between the copies and the input, and an information theoretic `quality'
measure given by the Holevo bound on the mutual information between the
density operators for the input and the copies reduced states have been discussed.
Although our copiers do not achieve the performance of other state-dependent
cloners known in the literature (because of the special choice of our
outputs), our results (which are new for the anti-cloning machine case) 
are still interesting as they show that the CPTP mapping `geometrical' methods 
are simpler and more direct than the study of the several constraints inherent
to the extended Hilbert space approaches. 
It would be interesting to compare our results on cloning with the
conditions discussed in Refs. \cite{cerf1} for Pauli cloning
machines, which seem to derive, albeit using a different analysis,
an intriguely similar geometric picture.

Finally, It should be also mentioned that
the use of squeezers has been studied as another kind
of repeater for coherent states \cite{hirota}.
In particular, it was shown that
by optimizing a cascade of squeezers
the communication performance of the coherent state channel
can be improved. This method is based on the unitary
transformation of the squeezer as a noiseless amplifier.
Therefore the state overlap between the signal states
is not changed, which meens that the Helstrom bound cannot
be improved.
However, considering homodyne detection (which is a practical detection
scheme with the present technology),
the improvement in the signal-to-noise ratio brought
by the cascade of the squeezers
will be very useful.
It would be an interesting problem to study quantum repeaters
combining our non-unitary repeater with the squeezer
repeater for a lossy channel with homodyne detection.

\acknowledgements
The authors acknowledge Prof. R. Jozsa for providing the original
motivation of this work and for crucial comments.
They also thank Dr. A. Chefles and Prof. O. Hirota for
valuable comments.

\newpage
\appendix
\section{~}

The solutions to the constraints (7) and (8) for the variables $p$ and
$q$ in terms of the  parameters $R, X$ and $ Y_0$ given by eq. (6)
can be summarized, after some lengthy but straightforward algebra, by
the geometrical pictures shown in Figs. A1-A7.
In particular, the allowed regions for the existence of the CPTP maps
between arbitrary mixed initial and final states are the shaded regions
in these figures, bounded by the following sets of curves:

a) the lines:
\bea
\hat q_{1~\pm}(\hat p)&\equiv &-{a_{\pm}\over X}(\sqrt{Y_0}+a_{\pm}\hat p),
\nonumber \\
\hat q_{2~\pm}(\hat p)&\equiv &{a_{\pm}\over X}(\sqrt{Y_0}-a_{\pm}\hat p),
\label{diamond}
\eea
(where we have defined $a_{\pm}\equiv 1\pm\sqrt{1+X}$) and 
\beq
\hat q_{3~\pm}\equiv \pm {\sqrt{Y_{0~X}}\over 2};
\label{pm}
\eeq

b) the conic (an ellipse for $R<Y_0/(1+X)$):
\beq
\Delta_{M}(\hat p, \hat q)\equiv Y_{0~X}({\hat p}^2+{\hat q}^2)+2Y_{2~X}
\hat p\hat q -RY_{1~X}.
\label{conic}
\eeq

The allowed regions for the variables $p$ and $q$ can then be classified
in different sets, defined by certain ranges for the values of the 
parameters $R, X$ and $Y_0$, and depending on the type of intersections
among the above curves and the global geometrical shape of the allowed
region itself.
In more details, we distinguish among the following sets of parameters:

\bea
1a)& Y_0>2;~0<X<Y_0-2;~0<R<1;
\\
1b)& \mathrm{max}(Y_0-2, 0)<X<Y_0-1;~0<R<{Y_0\over 2+X};
\\
1c)& Y_0-1<X<\sqrt{Y_0}(\sqrt{Y_0}+2);~0<R<{(Y_0-1)\over 1+X};
\\
1d)& X>\sqrt{Y_0}(\sqrt{Y_0}+2);~0<R<{\sqrt{Y_0}(\sqrt{Y_0}-1)\over
\sqrt{1+X}(\sqrt{1+X}-1)}
\label{a1}
\eea
(see Fig. 7) or:
\bea
2a)& Y_0>2;~0<X<Y_0-2;~0<R<1;
\\
2b)& \mathrm{max}(Y_0-2, 0)<X<Y_0-1;~{Y_0-1\over X}<R<{Y_0\over X};
\\
2c)& Y_0-1<X<4\sqrt{Y_0}(\sqrt{Y_0}-1);~R_0<R<{Y_0\over X}
\label{a2}
\eea
(see Fig. 8) or:
\bea
3a)& Y_0>2;~0<X<Y_0-2;~0<R<1;
\\
3b)& \mathrm{max}(Y_0-2, 0)<X<Y_0-1;~1<R<{Y_0-1\over X}
\label{a3}
\eea
(see Fig. 9) or:
\beq
4) X>Y_0-1;~{\sqrt{Y_0}(\sqrt{Y_0}-1)\over \sqrt{1+X}(\sqrt{1+X}-1)}<R<
{Y_0-1\over X}
\label{a4}
\eeq
(see Fig. 10) or:
\beq
5) Y_0-1<X<X_0;~{Y_0-1\over X}<R<{Y_0\over 1+X}
\label{a5}
\eeq
(see Fig. 11) or:
\bea
6)& X>4\sqrt{Y_0}(\sqrt{Y_0}-1);~{Y_0\over 1+X}<R<{Y_0\over X}
\label{a6}
\eea
(see Fig. 12) or, finally:
\bea
7)& X>X_0;~R_0<R<{Y_0\over 1+X}
\label{a7}
\eea
(see Fig. 13).
The values of $X_0$ and $R_0$ are to be determined numerically.
For instance, in the case $Y_0=4$ we obtain $X_0\simeq 20$ and 
$R_0(X)=[3X^2+4(X-2)\sqrt{1+X}-8]/X^3$.

\newpage
\begin{figure}
\begin{center}
\includegraphics[width=0.45\textwidth, height=6cm]{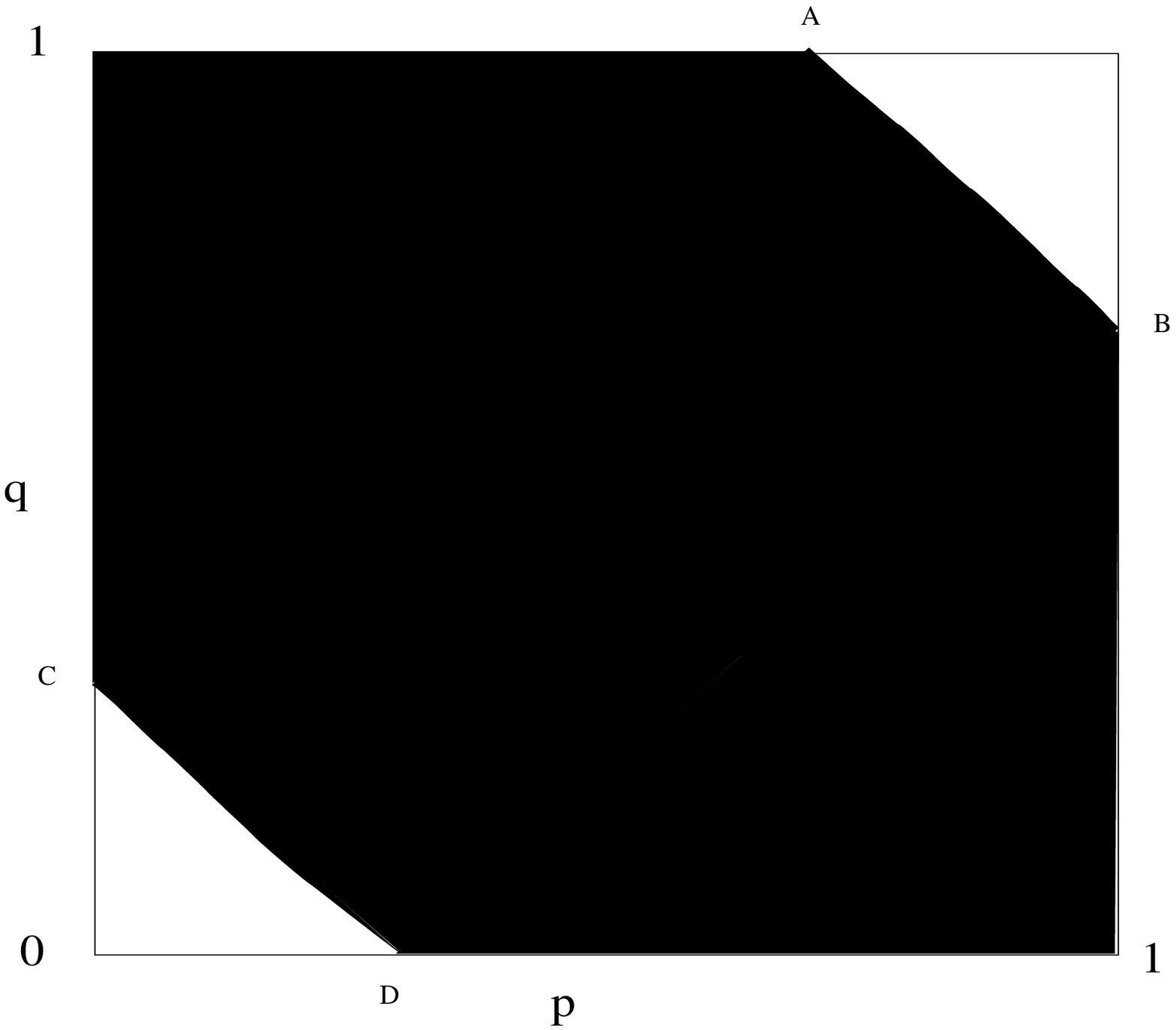}
\end{center}
\caption{\label{Fig. A1}
The allowed $(p, q)$ region (shaded area) for the existence of a CPTP mapping
between two input mixed states and two output mixed states for
the set of parameters: $Y_0=4,~~X=1,~~R=0.5$ (case 1)).
Points $A, B, C, D$ represent, in the order, the intersections of the 
ellipse \rf{conic} with the boundaries $q=1, p=1, p=0$ and $q=0$.}
\end{figure}
\begin{figure}
\begin{center}
\includegraphics[width=0.45\textwidth, height=6cm]{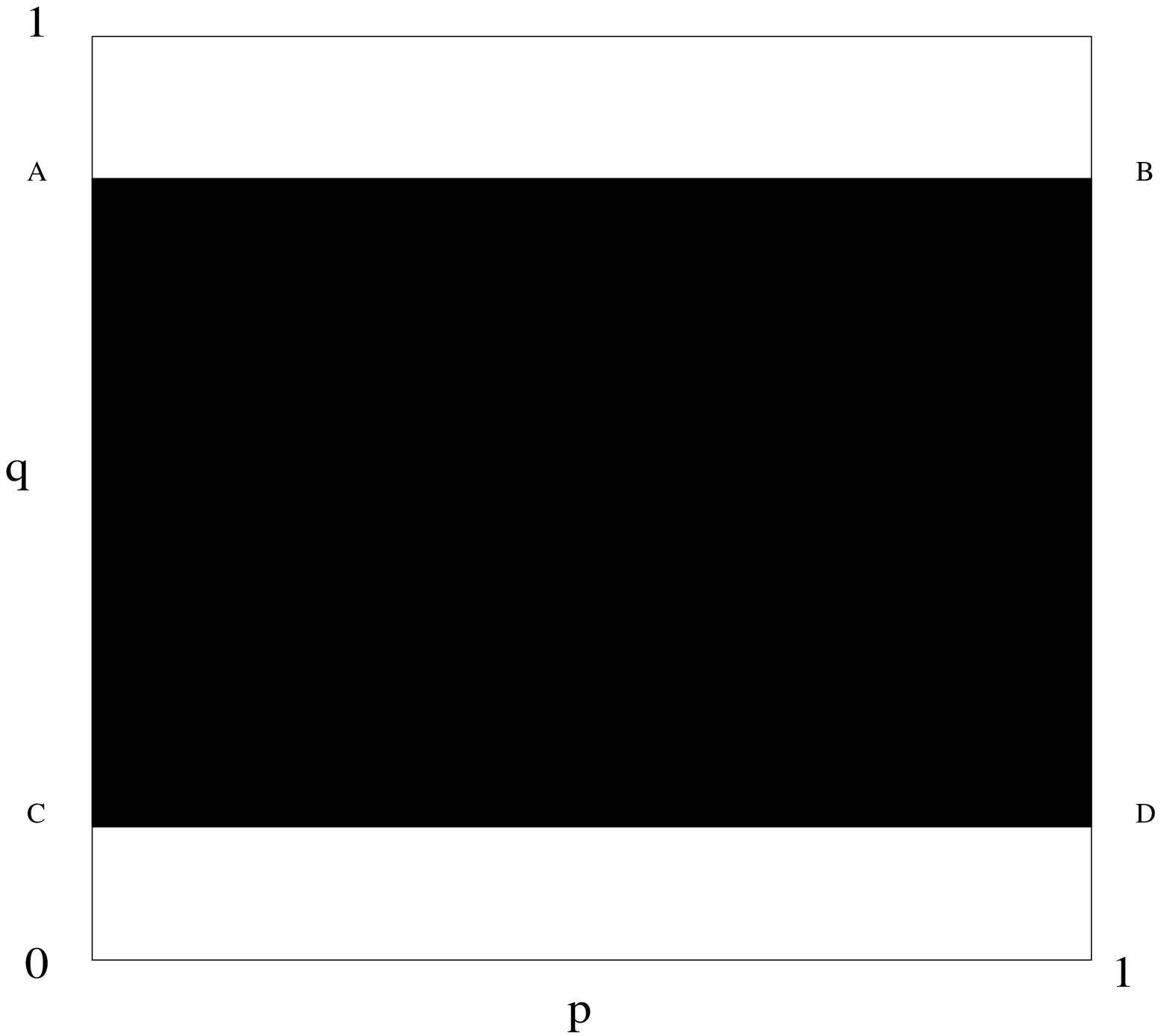}
\end{center}
\caption{\label{Fig. A2}
The allowed $(p, q)$ region (shaded area) for the existence of a CPTP mapping
between two input mixed states and two output mixed states for
the set of parameters: $Y_0=4,~~X=1,~~R=3.5$ (case 2)).
Points $A, B, C, D$ represent, in the order, the intersections of the 
lines $\hat q_{3~\pm}$ with the boundaries $p=0, p=1, p=0$ and $p=1$.}
\end{figure}
\begin{figure}
\begin{center}
\includegraphics[width=0.45\textwidth, height=6cm]{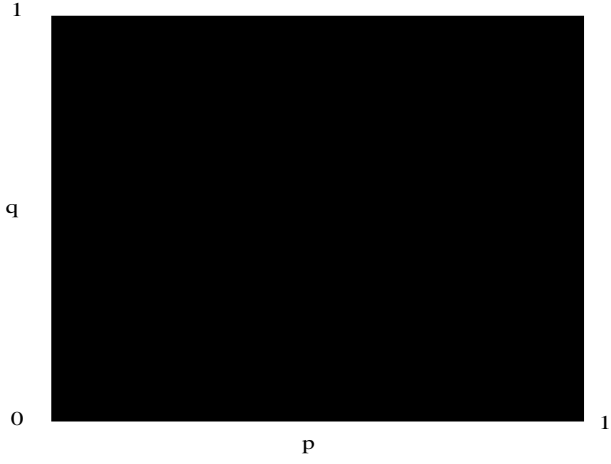}
\end{center}
\caption{\label{Fig. A3}
The allowed $(p, q)$ region (shaded area) for the existence of a CPTP mapping
between two input mixed states and two output mixed states for
the set of parameters: $Y_0=4,~~X=1,~~R=2.5$ (case 3)).}
\end{figure}
\begin{figure}
\begin{center}
\includegraphics[width=0.45\textwidth, height=6cm]{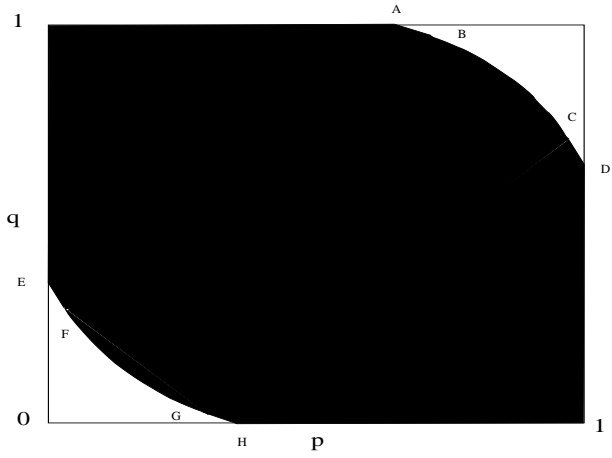}
\end{center}
\caption{\label{Fig. A4}
The allowed $(p, q)$ region (shaded area) for the existence of a CPTP mapping
between two input mixed states and two output mixed states for
the set of parameters: $Y_0=4,~~X=5.5,~~R=0.52$ (case 4)).
The lines \rf{diamond} intersect with the boundaries $q=1, p=1, p=0$ and $q=0$
at the points $A, D, E, H$, and with ellipse \rf{conic} at the points
$B, C, F, G$.}
\end{figure}
\begin{figure}
\begin{center}
\includegraphics[width=0.45\textwidth, height=6cm]{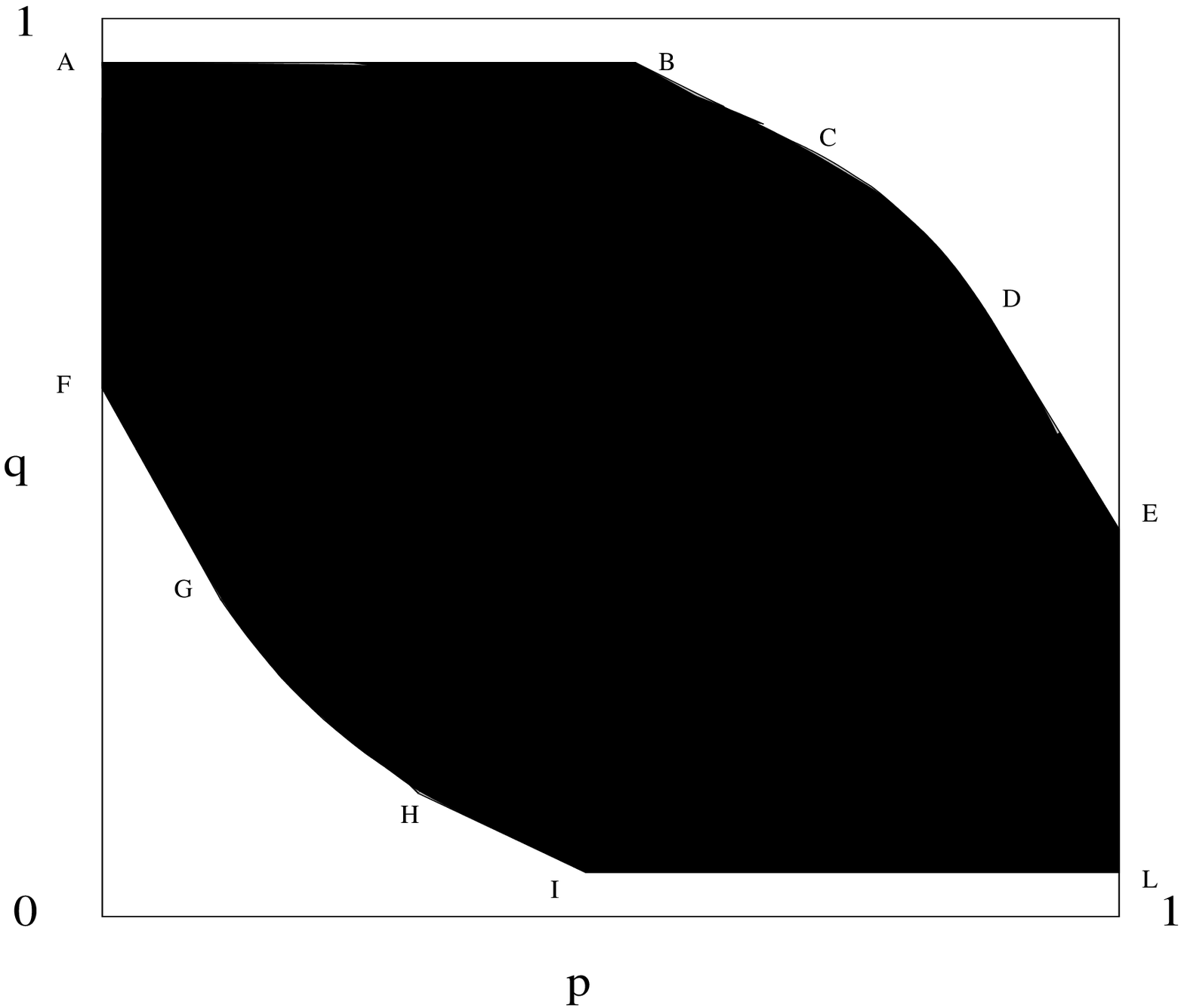}
\end{center}
\caption{\label{Fig. A5}
The allowed $(p, q)$ region (shaded area) for the existence of a CPTP mapping
between two input mixed states and two output mixed states for
the set of parameters: $Y_0=4,~~X=10,~~R=0.32$ (case 5)).
The lines \rf{diamond} intersect with the the horizontal lines $\hat q_{3~\pm}$
at the points $B, I$ and with the boundaries $p=0, p=1$ at $F, E$; the
horizontal lines $\hat q_{3~\pm}$ intersect with the boundaries $p=0, p=1$
at $A, L$; the ellipse \rf{conic} intersects with the lines \rf{diamond} 
at the points $C, D, G, H$.}
\end{figure}
\begin{figure}
\begin{center}
\includegraphics[width=0.45\textwidth, height=6cm]{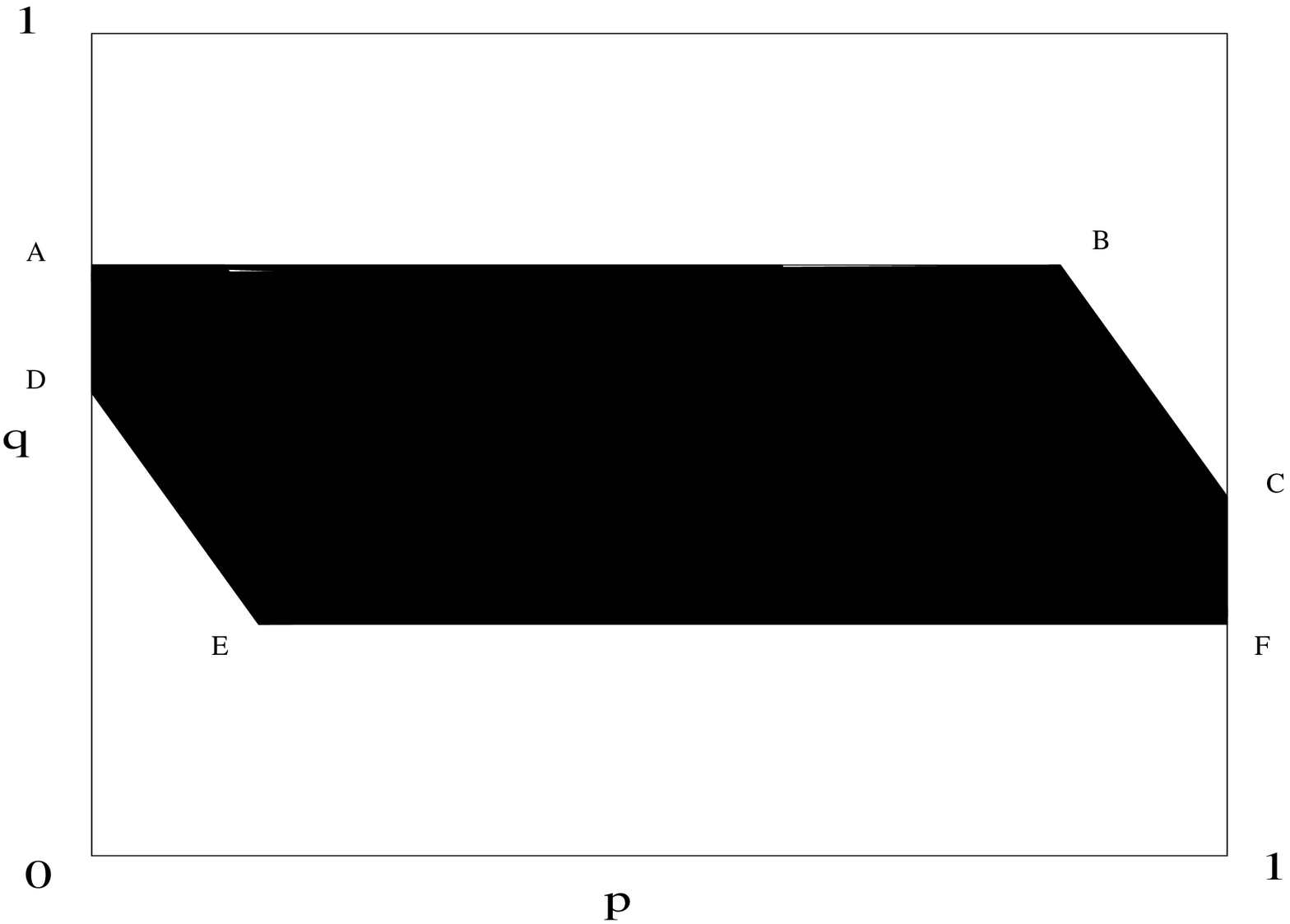}
\end{center}
\caption{\label{Fig. A6}
The allowed $(p, q)$ region (shaded area) for the existence of a CPTP mapping
between two input mixed states and two output mixed states for
the set of parameters: $Y_0=4,~~X=10,~~R=0.38$ (case 6)).
The lines \rf{diamond} intersect with the the horizontal lines $\hat q_{3~\pm}$
at the points $B, E$ and with the boundaries $p=0, p=1$ at $D, C$; the
horizontal lines $\hat q_{3~\pm}$ intersect with the boundaries $p=0, p=1$
at $A, F$.}
\end{figure}
\begin{figure}
\begin{center}
\includegraphics[width=0.45\textwidth, height=6cm]{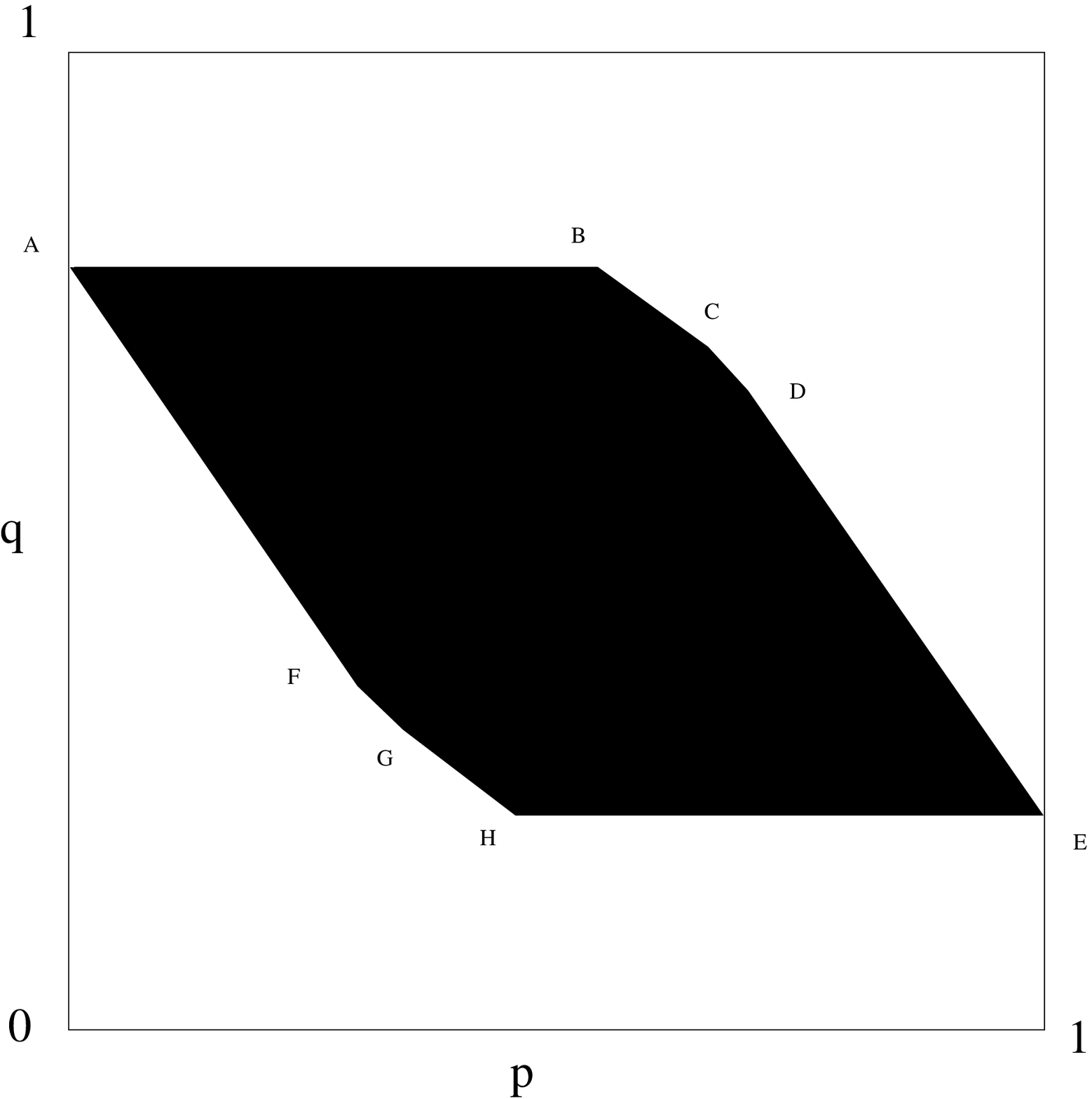}
\end{center}
\caption{\label{Fig. A7}
The allowed $(p, q)$ region (shaded area) for the existence of a CPTP mapping
between two input mixed states and two output mixed states for
the set of parameters: $Y_0=4,~~X=30,~~R=0.123$ (case 7)).
The lines \rf{diamond} intersect with the the horizontal lines $\hat q_{3~\pm}$
at the points $B, H$, with the boundaries $p=0, p=1$ at $A, E$, and
with the ellipse at the points $C, D, F, G$.}
\end{figure}


\begin{references}

\bibitem{albertiuhlmann}
P.M. Alberti and A. Uhlmann, Rep. Math. Phys. 18, 163 (1980).

\bibitem{csj}
A. Carlini and M. Sasaki, in preparation, to be submitted to Phys.
Rev. A (2002).

\bibitem{uhlmann}
A. Uhlmann, Rep. Math. Phys. 9, 273 (1976).

\bibitem{helstrom}
C.W. Helstrom, Quantum Detection and Estimation Theory (Academic
Press, New York, 1976).

\bibitem{sasaki}
M. Sasaki, T.S. Usuda, O. Hirota and A.S. Holevo, 
Phys. Rev. A53, 1273 (1996).

\bibitem{Cochrane99}
P. T. Cochrane, G. J. Milburn, and W. J. Munro, 
Phys. Rev. A{\bf59}, 2631 (1999).

\bibitem{wootterszurek}
W.H. Wootters and W.H. Zurek, Nature 299, 802 (1982);
D. Dieks, Phys. Lett. A126, 303 (1988);
H. Barnum, C. Caves, C. Fuchs, R. Jozsa and B. Schumacher, 
Phys. Rev. Lett. 76, 2818 (1996).

\bibitem{fanmatsumotowadati}
H. Fan, K. Matsumoto and M. Wadati, Phys. Rev. A64, 064301 (2001);
A. Chefles and S.M. Barnett, J. Phys. A31, 10097 (1998).


\bibitem{bruss1}
D. Bru\ss, D.P. DiVincenzo, A.K. Ekert, C.A. Fuchs, C. Macchiavello and
J. Smolin, Phys. Rev. A57 , 2368 (1998).

\bibitem{chefles2}
A.  Chefles and S.M.  Barnett, Phys.  Rev.  A60, 136 (1999).

\bibitem{buzekhillery1}
V. Buzek and M. Hillery, Phys. Rev. A54, 1844 (1996);
M. Hillery and V. Buzek, Phys. Rev. A56, 1212 (1997);
N. Gisin and B. Huttner, Phys. Lett. A228, 13 (1997);
D. Bru\ss ~ and C. Macchiavello, e-print archive quant-ph/0110099.

\bibitem{deuar}
P. Deuar and W.J. Munro, Phys. Rev. A61, 062304 (2000).

\bibitem{footnote}
To compare the results 
of Refs. \cite{bruss1} (input overlap $S\equiv \sin 2\theta$) and 
\cite{deuar} (input overlap $f$) discussed above, we note that  
$Z\equiv \sqrt{S}\equiv \sqrt{f}$ and, consequently, 
$\phi=-2\theta +\pi/2$ ($\phi\in [0, \pi/2]$.

\bibitem{songhardy}
D.D. Song and L. Hardy, e-print archive quant-ph/0001105.

\bibitem{gisinpopescu}
C.H. Bennett and S.J. Wiesner, Phys. Rev. Lett. 69, 2881 (1992);
N.J. Cerf and C. Adami, Phys. Rev. Lett. 79, 5194 (1997);
N. Gisin and S. Popescu, Phys. Rev. Lett. 83, 432 (1999);
V. Buzek, M. Hillery and R.F. Werner, Phys. Rev. A60, R2626 (1999);
N.J. Cerf and S. Iblisdir, Phys. Rev. Lett. 87, 247903 (2000).

\bibitem{cerf1}
N.J. Cerf, Phys. Rev. Lett. 84, 4497 (2000);
N.J. Cerf, e-print archive quant-ph/9805024.

\bibitem{hirota}
O. Hirota, Squeezed Light (Elsevier, Amsterdam, 1992).

\end{references}
\end{document}